%% file: MAIN_Arxiv.tex
\documentclass{article}

\usepackage{PRIMEarxiv}

\usepackage[utf8]{inputenc} % allow utf-8 input
\usepackage[T1]{fontenc}    % use 8-bit T1 fonts
\usepackage{url}            % simple URL typesetting
\usepackage{booktabs}       % professional-quality tables
\usepackage{amsfonts}       % blackboard math symbols
\usepackage{nicefrac}       % compact symbols for 1/2, etc.
\usepackage{microtype}      % microtypography
\usepackage{lipsum}
\usepackage{fancyhdr}       % header
\usepackage{graphicx}       % graphics
\usepackage{natbib}
\usepackage[colorlinks=true, linkcolor=blue, citecolor=blue, urlcolor=blue]{hyperref}

\usepackage{iftex}
\usepackage{nopageno}
\usepackage{geometry}
\usepackage{subcaption}
\usepackage{parskip}
\usepackage{amsmath,amssymb}%
\usepackage{cuted}%
\usepackage{xurl}
\usepackage[all]{nowidow}
\usepackage{calc}%
\usepackage{xcolor}%
\usepackage{float}%
\usepackage{array}%
\usepackage{tabularray}
\usepackage{tikzpagenodes}
\usepackage{tikz}
\usepackage[figuresright]{rotating}%
\usepackage{afterpage}%
\usepackage{enumitem}%
\usepackage{soul}
\usepackage{newtxtext}

\input{utils/defined_colors_and_commands}

\graphicspath{{media/}}     % organize your images and other figures under media/ folder

\title{Genetic Optimization of a Software-Defined GNSS Receiver}

\author{
  Laura Train \\
  Escuela T\'ecnica Superior de Ingenier\'ia Aeron\'autica y del Espacio (ETSIAE)\\
  Universidad Politécnica de Madrid\\
  28040 Madrid, Spain\\
  \texttt{l.train@alumnos.upm.es} \\
   \And
  Rodrigo Castellanos \\
  Department of Aerospace Engineering \\
  Universidad Carlos III de Madrid \\
  28911 Madrid, Spain \\
  \texttt{rcastell@ing.uc3m.es} \\
  \And
  Miguel G\'omez-L\'opez \\
  Aerial Platforms Department\\
  National Institute for Aerospace Technology (INTA)\\
  28330 San Mart\'in de la Vega, Madrid, Spain\\
  and\\
  Escuela de Ingenier\'ia Industrial y Aeroespacial de Toledo\\
  Universidad de Castilla La Mancha\\
  45071 Toledo, Spain\\
  \texttt{gomezlma@inta.es} \\
}

\begin{document}
\maketitle
\begin{abstract}
Commercial off-the-shelf (COTS) Global Navigation Satellite System (GNSS) receivers face significant limitations under high-dynamic conditions, particularly in high-acceleration environments such as those experienced by launch vehicles. These performance degradations, often observed as discontinuities in the navigation solution, arise from the inability of traditional tracking loop bandwidths to cope with rapid variations in synchronization parameters.
Software-Defined Radio (SDR) receivers overcome these constraints by enabling flexible reconfiguration of tracking loops; however, manual tuning involves a complex, multidimensional search and seldom ensures optimal performance. This work introduces a genetic algorithm–based optimization framework that autonomously explores the receiver configuration space to determine optimal loop parameters for phase, frequency, and delay tracking. The approach is validated within an SDR environment using realistically simulated GPS L1 signals for three representative dynamic regimes—guided rocket flight, Low Earth Orbit (LEO) satellite, and static receiver—processed with the open-source GNSS-SDR architecture. Results demonstrate that evolutionary optimization enables SDR receivers to maintain robust and accurate Position, Velocity, and Time (PVT) solutions across diverse dynamic conditions. The optimized configurations yielded maximum position and velocity errors of approximately 
$6~\mathrm{m}$ and $0.08~\mathrm{m/s}$ for the static case, 
$12~\mathrm{m}$ and $2.5~\mathrm{m/s}$ for the rocket case, and $5~\mathrm{m}$ and $0.2~\mathrm{m/s}$ for the LEO case.
\end{abstract}

% keywords can be removed

%----------------------------------------

\section{Introduction}
Global Navigation Satellite System (GNSS) receivers have become integral components within the on-board guidance computers for LEO satellites \citep{gnss-leo}, space launchers \citep{gnss-launchers} and unmanned aerial vehicles \citep{gnss-uav}. Their significance lies in their ability to  provide systems with global positioning information, offering substantial advantages over inertial navigation systems. 
Unlike inertial navigation, GNSS signals do not suffer from errors that accumulate over time \citep{gnss-vs-ins}, do not require sensor alignment or initial calibration, and are not as affected by mechanical sensitivities such as vibrations or thermal drift \citep{ins-calib}.  
Recent advances in satellite-based navigation techniques include the integration of GNSS with alternative sources like Signals of Opportunity (SoOp) \citep{leosoops}, LEO constellation signals \citep{leognss} \citep{iridium}, and security enhancements such as Galileo’s OSNMA authentication \citep{osnma}. While these developments enhance GNSS robustness and availability, a concurrent shift is occurring in the design and testing of GNSS receivers, enabled by the adoption of software-defined radio (SDR) technology. SDR-based GNSS receivers perform all signal processing tasks in software, enabling flexible reconfiguration of internal architectures, systematic evaluation of algorithmic modifications, and direct control over low-level parameters \citep{sdr-foundation}. This level of adaptability facilitates advanced experimental research and automated parameter tuning, particularly in dynamic environments where COTS receivers exhibit significant limitations.

GNSS PVT solution computation is especially challenging for vehicles under high dynamics, due to the Doppler effect in the carrier and code frequency of the received signal generated by the relative dynamics between transmitter and receiver \citep{pll-assisted-fll}. GNSS synchronization is required for computing PVT solution in any receiver. It is performed in two steps, acquisition and tracking. During acquisition, the code phase and Doppler shift are coarsely estimated. Tracking allows to estimate accurately the code phase, carrier phase and Doppler shift, and continuously adapts the receiver oscillator phase, delay and frequency to keep the synchronization of the GNSS receiver and the satellites. Phase-Locked Loops (PLL), Delay Locked Loops (DLL), and in some cases Frequency-Locked Loops (FLL) constitute the architecture of a GNSS receiver tracking loop \citep{robust-tracking}. The latter is included in high dynamic receivers due to its capability to handle rapid changes in the carrier frequency, which contributes to keep the synchronization during acceleration peaks \citep{pll-assisted-fll} \citep{fll_hydyn}. FLL estimates the difference in frequency between the local oscillator and the incoming satellite signal \citep{fll-justification}, thus monitoring the carrier dynamics. High dynamic environments typically require a larger tracking loop bandwidth \citep{robust-tracking}. To accommodate fast signal variations, tracking loop bandwidths must often be widened, especially in high dynamics scenarios. However, larger bandwidths increase susceptibility to thermal noise and oscillator instability \citep{roncaglio1}, needing a careful trade-off between the dynamic response and the accuracy. This trade-off is bounded by theoretical limits based on jitter and dynamic stress error analysis, as discussed in \citep{theoretical-limit}. Previous studies have proposed adaptive strategies to address this challenge, such as the steepest ascent method for dynamically adjusting tracking parameters under variable conditions \citep{steepest-ascent-method}.

This study considers three representative use cases that span a wide range of dynamic conditions relevant to GNSS receiver performance. The first is a high-acceleration sounding rocket scenario, characterized by extreme transient dynamics reaching up to 55 g during launch. The second involves a Low Earth Orbit (LEO) satellite platform, where high and sustained orbital velocities (above 7500 m/s) and varying satellite geometry impose significant synchronization challenges. Finally, a static user scenario is analyzed as a baseline, serving to contrast the behavior of tracking loops in the complete absence of the receiver's dynamics. Given the diversity of these operating conditions, selecting suitable tracking loop configurations becomes a nontrivial task. Each scenario imposes different requirements on the receiver architecture, particularly in terms of how the PLL, DLL, and FLL respond to Doppler shifts, frequency rates, and oscillator noise. Manually tuning the tracking loop is complex due to the highly multidimensional nature of the problem.
This research evaluates three test cases representing distinctly different dynamic conditions: a high-acceleration rocket launch scenario reaching up to 55 g, a LEO satellite case with relative velocities exceeding 7500 m/s, and a static user scenario serving as a baseline for comparison. The objective is to determine an optimal set of tracking loop parameters for each scenario, using an experimental recording-based testbench to simulate the signal environment, while the optimizer autonomously searches for the best loop configuration in each case. 

To address this, the tuning of GNSS tracking loops is formulated as an optimization problem. The objective is to automatically determine the best configuration of loop parameters that maximize navigation performance in each dynamic context. This is achieved through a Genetic Algorithm (GA) that systematically explores the parameter space, evaluating each candidate configuration using a realistic SDR-based testbench. The setup includes a GNSS signal simulator that generates GPS L1 signals corresponding to the dynamic scenarios, an RF front-end that digitizes the signal, and a software-defined GNSS receiver that processes it in real time. Each configuration is assessed through a cost function based on positioning and velocity accuracy, allowing the GA to converge toward an optimal receiver setup tailored to each of the three case studies.

 Previous studies have shown the effectiveness of Genetic Algorithms in aerospace applications, such as radar tracking in high dynamics \citep{radar-rockets}, aircraft dynamics identification \citep{aero-genetic}, adjusting gain-scheduling control systems \citep{ga-gain-scheduling}, or trajectory optimization in commercial aircraft \citep{atc-genetic}. The GA used in this research is called Hybrid Genetic Optimizer (HyGO). The use of such an algorithm represents a substantial improvement in solving complex multidimensional problems compared to conventional optimization scenarios. The software executing HyGO has undergone prior validation for solving turbulent flow control problems \citep{rodrigo-turbulent}, \citep{rodrigo-turbulent2}. 

The remainder of this paper is organized as follows. Section \ref{experimental-setup} presents the experimental testbench developed to obtain the raw GNSS data, and \ref{gnss-architecture} presents the architecture of the GNSS tracking loops. Section \ref{section-optimization} formulates the GNSS receiver tuning as an optimization problem, while \ref{ss:cost-function} examines the cost function used for the optimization, and \ref{ss:hygo} describes the details of the HyGO algorithm. Section \ref{test-cases-scenarios} presents the test cases and explains how the dynamics are derived. Section \ref{results} analyzes the behavior of the optimizer and discusses suitable tracking loop configurations. Section \ref{conclusion} concludes the paper.

\section{Experimental setup} \label{experimental-setup}
GNSS-SDR \citep{gnss-sdr} is an open-source software-defined radio framework capable of real-time GNSS processing, and serves as the foundation for all case studies presented in this work. Proven to operate reliably under high-dynamic conditions \citep{gnss-sdr-escribano}, it provides the flexibility and control necessary to access, configure, and analyze every stage of the signal processing chain —from satellite signal acquisition and tracking to navigation message decoding and observable computation needed for the PVT solution—. Running on a standard Ubuntu GNU/Linux machine, GNSS-SDR enables a level of visibility and customization that is not achievable with COTS receivers, making it a fundamental enabler of the methodology adopted in this research.

The experiment begins with the Trajectory Simulator, which generates the time evolution of the dynamic state of the platform: latitude ($\lambda$), longitude ($\phi$), altitude (h), velocity (v), acceleration (a) and jerk (j) for each of the test cases described in Section \ref{test-cases-scenarios}. This data is provided to the Spirent GSS9000 signal generator, which produces highly realistic simulated RF signal. The simulator allows for the selection of constellations, frequency bands, and signal power levels, as well as the ability to fix the simulation date to maintain a consistent skyplot. For this experiment, the GPS L1 band was chosen.  Signal power levels were set to approximately 45 dB-Hz for the LEO satellite, since no atmospheric attenuation or ground-level obstructions are present in space, resulting in stronger and more stable signal reception. In contrast, the rocket and static cases were configured with more realistic terrestrial values between 40 and 42 dB-Hz, accounting for atmospheric losses and local multipath effects.

The radio frequency output is then captured using a USRP-X310, an RF front-end for software-defined radio platforms that downconverts, filters, amplifies, and digitizes the GNSS L1 signal into raw IQ samples. These IQ samples are processed by GNSS-SDR. Its modular architecture allows for the full configuration of the receiver, including acquisition, tracking, decoding of navigation messages, and PVT computation parameters. The tracking loop bandwidths of the receiver, which are the main optimization variables in this study, are discussed in detail in Section \ref{gnss-architecture}. The result of this process is what \autoref{experimental-setup} denotes as the Realistic Trajectory, which is the navigation solution estimated from the simulated signal. This output includes the reconstructed PVT of the trajectory, and serves as the basis for evaluating tracking loop performance under each test condition. 

\begin{figure}[H]
    \centering
    \includegraphics[width=1\textwidth]{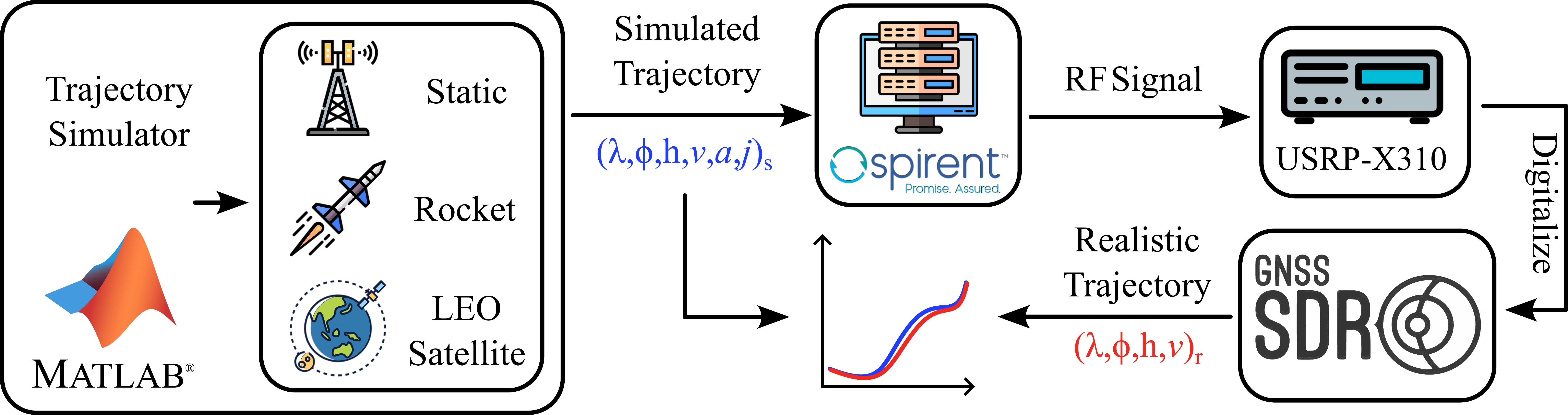}
    \caption{Experimental configuration}
\end{figure}

\subsection{GNSS architecture} \label{gnss-architecture}

\begin{figure}[H]
    \centering    \includegraphics[width=1\textwidth]{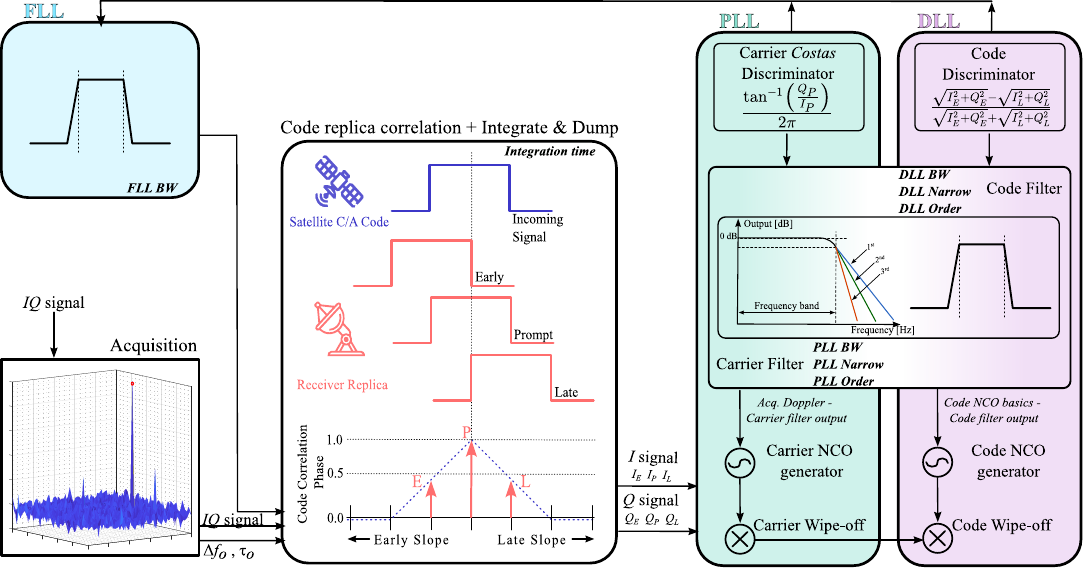}
    \caption{GNSS architecture used for the optimization loop, based on a FLL-assisted PLL.}
    \label{architecture}
\end{figure}

The GNSS receiver architecture adopted in this work is implemented within the GNSS-SDR framework and contains the DLL, PLL, and FLL loop blocks, as illustrated in \autoref{architecture}. Each of the loops relies on a discriminator function to extract the corresponding signal tracking error from the baseband correlator outputs. The DLL uses a code discriminator, typically implemented as the Early-minus-Late (E–L) envelope power difference, to estimate the code delay. The PLL employs a Costas discriminator, based on the arctangent of the ratio $Q/I$, to estimate the carrier phase error. The FLL uses a frequency discriminator derived from the differential phase evolution across consecutive integration intervals to estimate carrier frequency error. All discriminators operate on the in-phase ($I$) and quadrature ($Q$) components obtained after coherent integration and provide the necessary feedback to adjust the Numerically Controlled Oscillators (NCOs). Together, they enable accurate and continuous tracking of the fundamental signal parameters: code delay, carrier phase, and carrier frequency.

Each of these loops incorporates parameters that critically influence performance and robustness under dynamic conditions. These include the loop filter bandwidth and loop filter order. In high-dynamic environments, such as those involving accelerations beyond 4 g, large and rapidly changing Doppler shifts can make standard PLL/DLL-based architectures unstable, often preventing the receiver from maintaining continuous tracking and computing a reliable PVT solution. To address this limitation, the FLL block is introduced alongside the PLL \citep{steepest-ascent-method}. FLL directly estimates the carrier frequency and assists the PLL in tracking rapid frequency variations, improving lock robustness, reducing acquisition times, and helping to prevent false locks \citep{pll-assisted-fll}. While increasing loop bandwidths may be necessary to follow fast signal dynamics, such as frequency rate and jerk, this also introduces higher noise into the tracking estimates. Consequently, careful optimization of bandwidth and the inclusion of the FLL becomes essential to maintain signal lock and ensure navigation continuity in challenging motion scenarios. Similarly, the FLL order is selected to be one order less than the PLL \citep{theoretical-limit}.

Manual tuning of GNSS tracking loop parameters is the conventional approach used in most receivers to ensure signal synchronization. However, this method is often suboptimal in dynamic environments, where the performance of the loops becomes highly sensitive to multiple interacting variables. The tuning space is inherently multidimensional, and each parameter exerts a nonlinear influence on tracking performance, especially under varying dynamics such as acceleration and jerk. As such, relying on empirical adjustment can lead to poor tracking quality, increased noise, or even loss of lock. A systematic optimization strategy is therefore required to explore the parameter space comprehensively and derive configurations that ensure stable and accurate PVT across the different scenarios.

The optimization process therefore focuses on the following set of parameters:

\begin{itemize}
    \item \textbf{Integration time}: Determines the signal-to-noise ratio (SNR) of the correlator outputs by setting the duration over which incoming signal samples are accumulated.
    
    \item \textbf{PLL and DLL loop bandwidths (before bit synchronization):} Define the initial responsiveness of loops to frequency and delay changes. Wider bandwidths are needed to handle high-dynamic inputs, but increase noise contribution.

    \item \textbf{PLL and DLL narrow loop bandwidths (after bit synchronization):} Allow bandwidth adjustment once bit transitions no longer interfere with phase tracking, allowing narrower filters for improved stability in later tracking stages.
    
    \item \textbf{PLL and DLL loop filter orders}: Govern the agility and responsiveness of the loops to dynamic variations in the signal, such as frequency rate and jerk.
    
    \item \textbf{FLL loop bandwidth (before bit synchronization)}: Assists the PLL in handling rapid frequency variations by providing coarse frequency tracking, especially important in the pull-in process during high-dynamic scenarios to prevent loss of lock.
\end{itemize}

\section{Test Cases Definition} \label{test-cases-scenarios}
To assess the adaptability of GNSS tracking loops under different motion regimes, three representative test cases are defined: a LEO satellite, a high-acceleration sounding rocket, and a completely static receiver. These scenarios are selected to span a wide spectrum of dynamic conditions: from extreme accelerations and jerk, to high relative velocities, and finally to a baseline case with no movement. Each test case is modeled to accurately replicate its physical environment and associated dynamics by formulating the governing equations of motion and integrating them numerically to obtain the Ground Truth (GT) trajectory. This GT solution serves as the reference input for GNSS signal generation and as the baseline for evaluating receiver optimization performance.
    
\subsection{LEO satellite}
The LEO platform selected for this study is the International Space Station (ISS). The motion is governed by the two-body equation, where the gravity attraction is the dominant force. The main perturbation forces are Earth's oblateness ($J_2$ effect) and the atmospheric drag. The modeling of the perturbation effects is described.
\begin{itemize}
    \item Due to the Earth's equatorial bulge, the gravitational field deviates from perfect sphericity. This perturbation introduces secular variations in orbital elements such as the right ascension of the ascending node and the argument of perigee. The $J_2$ acceleration is modeled as:
    \begin{equation}
        \mathbf{a}_{J_2} = \frac{3}{2} J_2 \frac{\mu R_e^2}{r^5} 
\begin{bmatrix}
x \left(1 - 5 \frac{z^2}{r^2} \right) \\
y \left(1 - 5 \frac{z^2}{r^2} \right) \\
z \left(3 - 5 \frac{z^2}{r^2} \right)
\end{bmatrix}
    \end{equation}
    Where $J_2$ is the second zonal harmonic coefficient, $R_e$ is the Earth's equatorial radius, $\mu$ is the Earth’s standard gravitational parameter, $r$ is the geocentric distance of the satellite, and $x$, $y$, and $z$ are the components of the satellite's position in the Earth-Centered Inertial (ECI) reference frame.
    \item At LEO altitudes, residual atmospheric particles exert a significant drag force on satellites, leading to gradual orbital decay. This force depends on atmospheric density, the satellite's shape and size, and its velocity relative to the atmosphere. It is typically modeled as:
    \begin{equation}
        \mathbf{a}_{\text{drag}} = -\frac{1}{2} \frac{C_D A}{m} \, \rho \, v_{\text{rel}} \, \mathbf{v}_{\text{rel}}
    \end{equation}
    Where $C_D$ is the drag coefficient, $A$ is the cross-sectional area, $m$ is the mass of the satellite, $\rho$ is the atmospheric density, $\mathbf{v}_{\text{rel}}$ is the velocity relative to the rotating Earth reference frame.
\end{itemize}

The optimization test case is a 10-minute segment of the ISS orbit simulated using the physical and orbital parameters listed in \autoref{Table:QB50_ISS_orbits}. The second column in \autoref{ground_truth_traj} contains the details of the dynamics of the LEO satellite. The signal acquisition is performed in-orbit, and for this reason the maximum Doppler rate is set to 40 kHz.

\begin{table}[H]
    \centering
    \begin{tabular}{l c}
    \hline
    \textbf{Parameter} & \textbf{ISS}\\
    \hline \hline
    Altitude  & 417 km \\
    Eccentricity  & 0.0004413 \\
    Inclination & 51.6479\textdegree\\
    Argument of perigee  & 109.5258\textdegree\\
    RAAN & 103.0788\textdegree \\
    True anomaly  & 31.6858\textdegree \\
    \hline
    \end{tabular}
    \caption{ISS Orbital elements}
    \label{Table:QB50_ISS_orbits}
\end{table}

In our case analysis, the output of the dynamic model was used as the GT trajectory for validation purposes. However, in a real observation scenario, relying solely on the dynamic model is insufficient to determine the true trajectory of the satellite over time. This is due to the accumulation of errors caused by uncertainties in initial conditions, unmodeled perturbations, and inaccuracies in force models. To correct and constrain the propagation of the orbit, external observational data are essential. Embedding a GNSS receiver onboard the satellite enables the acquisition of continuous and precise PVT measurements. These observations, when combined with the dynamic model through estimation techniques (i.e Extended Kalman Filter), allow for a highly accurate reconstruction of the satellite’s orbit. In the LEO case, the Doppler shift is substantial, but its rate of change is smooth and continuous due to orbital mechanics. The jerk sensitivity is low in orbit, but the non-negligible acceleration necessitates a higher loop order without requiring significantly higher bandwidth \citep{theoretical-limit}.

\subsection{Sounding Rocket}
The rocket modeled in this study is a solid-propulsion sounding rocket, whose translational dynamics are divided into two main phases: powered flight and coasting \citep{sounding-rockets}. During the powered phase, the dominant force is thrust, whereas gravitational and aerodynamic contributions are secondary. This results in a high-acceleration regime with rapidly increasing velocity. Once the propellant is depleted, the vehicle enters the coasting phase, characterized by free-flight dynamics and a sudden deceleration. The motion of the rocket is governed by the following:
\begin{equation}
    m \cdot \ddot{\mathbf{x}} = \mathbf{F}_L + \mathbf{F}_A + \mathbf{W} + \mathbf{T}
\end{equation}

Where $\mathbf{F_L}$ and $\mathbf{F_A}$ are the lateral and axial aerodynamic forces, $\mathbf{W}$ is the weight vector,  $\mathbf{T}$ is the thrust vector produced by the solid-rocket motor, and $m$ is the mass. 

The dynamics were simulated using a 5-DOF mission simulator, validated according to the STANAG 4355 NSO 2017 standard \citep{5dof}. The 5-DOF framework assumes the axial symmetry of the vehicle and simplifies rotational dynamics by treating angular momentum separately in the axial and normal directions. Compared to a full 6-DOF model, this simplification reduces computational cost while preserving essential trajectory behavior. The rocket follows a ballistic trajectory under extreme dynamic conditions. 

\begin{figure}[h]
    \centering
    \includegraphics[width=0.9\textwidth]{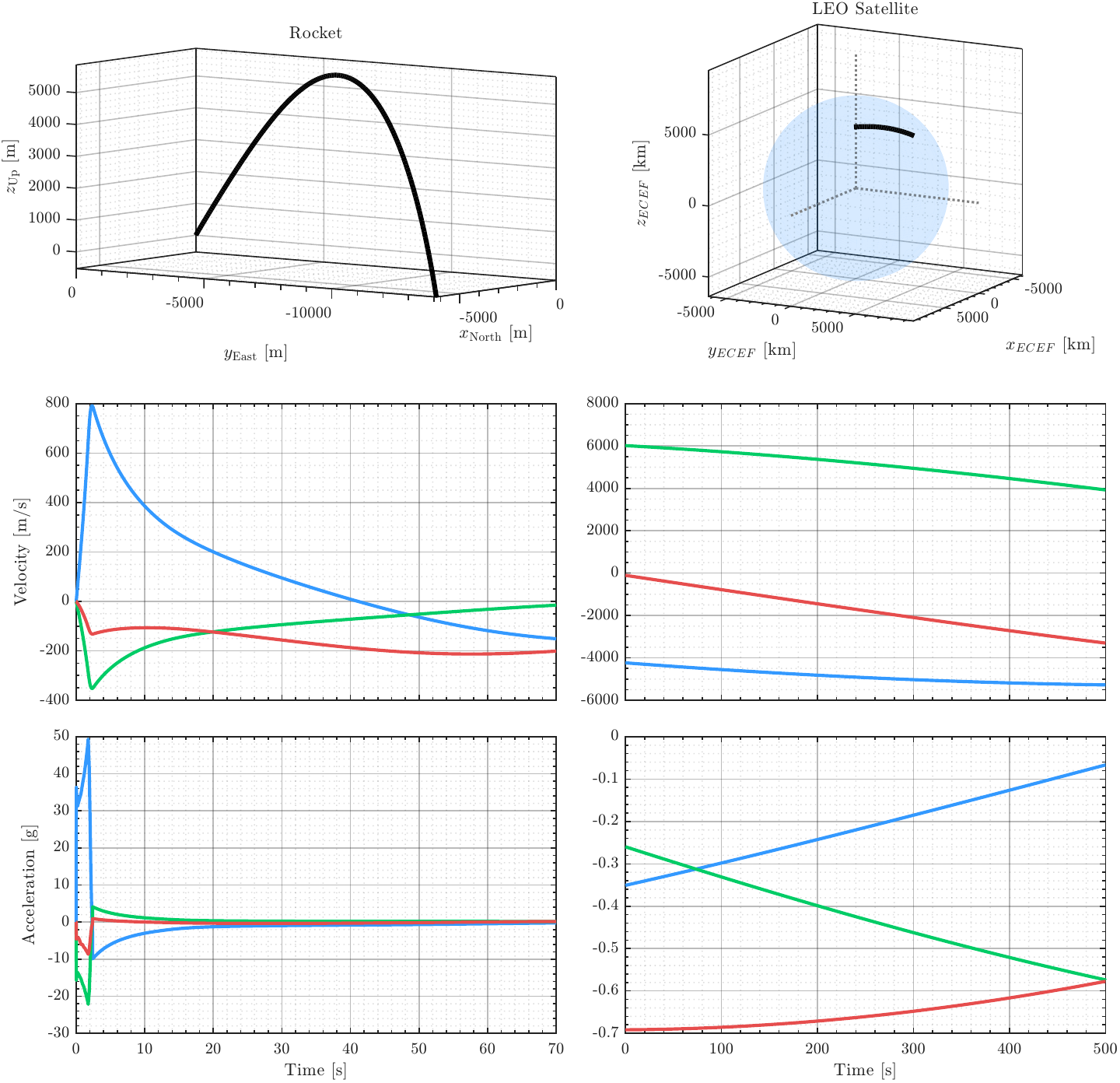}
    \caption{Ground truth. \textbf{First row}: trajectory; \textbf{Second row}: velocity; \textbf{Third row}: acceleration. The axes are represented such that (\textcolor[rgb]{0.2, 0.6, 1.0}{\rule[0.5ex]{0.5cm}{1pt}}) is x-axis, (\textcolor[rgb]{0.0, 0.8, 0.4}{\rule[0.5ex]{0.5cm}{1pt}}) is y-axis, (\textcolor[rgb]{0.9, 0.3, 0.3}{\rule[0.5ex]{0.5cm}{1pt}}) is z-axis in ECEF reference frame.}
    \label{ground_truth_traj}
\end{figure}

The test case for the launch scenario consists of a 70-second rocket trajectory, simulated under a launch angle of 45 degrees. The flight reaches an apogee of 5885 meters, with a maximum velocity of 875 m/s and a peak acceleration of 40 g during the powered phase. The trajectory and dynamic profile are illustrated in the first column of \autoref{ground_truth_traj}. The rocket performs the acquisition process statically, 2 minutes before the actual launch. Since the goal is to study only the high dynamic trajectory, the static phase is not included in the optimization problem. Further challenges are expected in rockets that need to acquire the signal once they are launched. This is covered in the LEO satellite scenario.

GNSS also plays a critical role in launch scenarios, although its performance is challenged by the extreme dynamics involved. Most commercial GNSS receivers are not designed to operate reliably beyond 4 g of acceleration. In this test case, although acceleration levels exceed this threshold for only about 1 second, the receiver may lose lock and require several seconds to reacquire the signal. During this recovery period, navigation relies solely on inertial systems, whose performance can degrade significantly under high acceleration conditions due to sensor saturation and integration drift. Therefore, it is essential to have a receiver capable of maintaining tracking regardless of acceleration levels. This scenario introduced the most challenging dynamics, with high velocity, acceleration, and jerk. This usually requires wide bandwidths and high-order PLLs, often assisted by FLLs \citep{theoretical-limit}. 

\subsection{Static}
A static scenario was also evaluated to analyze the behavior of the GNSS tracking loops under minimal dynamic stress, with zero velocity and acceleration. Due to negligible dynamics, this case served as a baseline for noise-limited performance. A 3-minute segment in fully static conditions was simulated to assess this behavior. In addition, this setup allows to assess receiver stability over time, as any fluctuations in the estimated position can be directly attributed to internal noise, clock drift, or local multipath. 

%------------------------------------------------------------
%----------------------- HYGO -------------------------------
%------------------------------------------------------------
\section{GNSS tracking loop tuning as an optimization problem} \label{section-optimization}
Building on the experimental setup and benchmark scenarios described above, the GNSS receiver optimization problem is formulated as an open-loop, model-free optimization task over the space of possible receiver design variables. The objective is to identify the optimal receiver configuration (specifically, the tracking loop parameters) that achieves the best possible navigation performance in each dynamic regime. Inherently, the objective landscape is noisy, nonconvex, and potentially discontinuous, which poses significant challenges for traditional deterministic optimization approaches. Algorithms such as gradient descent, conjugate gradient, particle swarm optimization, differential evolution, or sequential quadratic programming have been explored in related GNSS optimization contexts \citep{optimization-methods,steepest-ascent-method}. However, many of these are best suited for smooth and differentiable cost functions and are prone to entrapment in local minima \citep{descent-vs-ga,particle-deep}. To robustly address the high dimensionality and complexity of the parameter search space, this work employs a genetic algorithm (specifically, the Hybrid Genetic Optimizer, HyGO) to enhance convergence, systematically explore global minima, and mitigate sensitivity to noise and discontinuities \cite{Robledo2025Hygo}.

Accordingly, the receiver optimization is formulated as an open-loop, model-free parametric optimization in an eight-dimensional design space, where each point in this space corresponds to a distinct set of GNSS receiver configuration parameters. Let the \textit{settings} vector be
$\boldsymbol{\theta}\in\mathbb{R}^{N_p}$, where $N_p=8$ represents the number of design variables (i.e., tracking loop and integration control parameters). The receiver, denoted by $\mathbf{K}(\mathbf{s}(t);\boldsymbol{\theta})$, processes the measurement sequence $\mathbf{s}(t)$ according to both $\boldsymbol{\theta}$ and $\Theta$, which encapsulate the internal and external conditions not subject to optimization in this study (e.g., front-end characteristics, simulation boundary conditions, or other fixed algorithmic settings).

The optimization objective is to select the settings vector $\theta$ so that the receiver configuration $\mathbf{K}(\mathbf{s}(t);\boldsymbol{\theta})$ minimizes a scalar cost $J$, which aggregates performance metrics (such as positioning and velocity accuracy) and potential penalty terms. The construction of $J$ is detailed in Subsection \ref{ss:cost-function}. This formulation is expressed as:
\begin{equation}
    \mathbf{K}^* = \arg \min_{\mathbf{K} \in \mathcal{K}} J(\mathbf{K}(\mathbf{s}(t);\boldsymbol{\theta});\Theta)
\end{equation}
where $\mathcal{K}$ defines the admissible domain for the receiver settings, bounded by both physical and algorithmic parameter limits, and $\Theta$ comprises all parameters held constant during the optimization procedure.

The receiver configuration encompasses up to eight control parameters: the integration time, PLL and DLL pull-in and narrow bandwidths, respective filter orders, and FLL bandwidth. The integration-time controller is exclusive to the static scenario since, in high-dynamics (rocket, LEO) cases, it must be short due to rapid signal variations and is therefore fixed at $1\mathrm{ms}$. The selectable ranges for each parameter, as detailed in \autoref{params-resolution}, are set wide enough to ensure the optimizer is not unduly constrained by regime-specific dynamics, thereby supporting generalizable comparison and robust assessment across all studied scenarios.

\begin{table} 
    \centering
    \begin{tabular}{cccccccc}
        \hline
        \multicolumn{1}{c} {\textbf{\begin{tabular}[c]{@{}c@{}}Integration \\ Time {[}ms{]}\end{tabular}}} & \textbf{\begin{tabular}[c]{@{}c@{}}PLL BW\\ {[}Hz{]}\end{tabular}} & \textbf{\begin{tabular}[c]{@{}c@{}}PLL BW \\ narrow \\ (\% PLL)\end{tabular}} & \textbf{\begin{tabular}[c]{@{}c@{}}PLL \\ filter order\end{tabular}} & \textbf{\begin{tabular}[c]{@{}c@{}}DLL BW\\ {[}Hz{]}\end{tabular}} & \textbf{\begin{tabular}[c]{@{}c@{}}DLL BW \\ narrow \\ (\% DLL)\end{tabular}} & \textbf{\begin{tabular}[c]{@{}c@{}}DLL \\ filter order\end{tabular}} & \textbf{\begin{tabular}[c]{@{}c@{}}FLL BW\\ {[}Hz{]}\end{tabular}} \\ \hline \hline 1 - 20 & 5 - 80 & 0 - 100 & {2,3} & 1 - 50& 0 - 100 & {1,2,3} & 1 - 50 \\ \hline
    \end{tabular}
    \caption{Controllers range for the optimization. The PLL/DLL $\textit{narrow}$ bandwidths are expressed as percentages of their respective $\textit{pull-in}$ bandwidths, where 0\% represents the minimum PLL/DLL $\textit{pull-in}$ value and 100\% corresponds to the actual $\textit{pull-in}$ bandwidth. This ensures that no configuration has a $\textit{narrow}$ bandwidth larger than its corresponding $\textit{pull-in}$ value. For example, in a configuration with a 50 Hz PLL \textit{pull-in} bandwidth, a 0\% PLL \textit{narrow} bandwidth corresponds to 5 Hz, while 100\% corresponds to 50 Hz. }
    \label{params-resolution}
\end{table}

%------------------ Cost Function-------------------------
\subsection{Cost function} \label{ss:cost-function}
The objective of the optimization is to minimize the three-dimensional position and velocity errors of the GNSS navigation solution. Ground truth (GT) trajectories and velocities are provided by the dynamic simulator, and the optimization evaluates the time-integrated squared error between the GT and the navigation outputs obtained via the receiver configuration $\mathbf{K}(\mathbf{s}(t);\boldsymbol{\theta})$.  Specifically, the position and velocity cost functions are expressed as the mean squared errors over the trajectory duration:
\begin{equation}
    J_{pos} = \frac{\int_{t_0}^{t_f} |\mathbf{r_{GT}} - \mathbf{r_{NAV}}|^2 dt}{tf-t0} \qquad \text{and} \qquad
    J_{vel} = \frac{\int_{t_0}^{t_f} |\mathbf{v_{GT}} - \mathbf{v_{NAV}}|^2 dt}{tf-t0}
\end{equation}
where $\mathbf{r}$ and $\mathbf{v}$ denote position and velocity vectors, respectively. Both the navigation solution and GT trajectories are time-aligned using GPS time stamps and sampled at $20\,\mathrm{Hz}$ to enable precise point-wise comparison. This integral cost formulation ensures that solutions exhibiting transient discontinuities or sustained errors are penalized, thus encouraging stable and accurate tracking across the entire trajectory.

This problem is inherently multi-objective, as position and velocity errors represent distinct metrics of receiver performance. However, the optimization algorithm used here, HyGO, is single-objective, necessitating scalarization of the multi-objective costs into a unified scalar cost for evaluation and comparison during optimization. This scalarization is achieved by weighting and summing the position and velocity costs.

A primary challenge in combining these costs is their differing physical units and magnitudes (position errors are measured in squared meters, while velocity errors are in squared meters per second squared), making direct summation inappropriate and potentially biasing the optimization towards the cost with larger numerical values. To address this and guarantee a fair treatment where both components are equally weighted, a \textit{min-max normalization} is applied to both $J_{pos}$ and $J_{pos}$, scaling their values into the range $[0,1]$. This normalization can be viewed as a data-driven rescaling that avoids heuristic or arbitrary assignment of scalar weights, thus ensuring that the optimizer explores the parameter space with balanced sensitivity to both position and velocity performance. The overall normalized and scalarized cost function thus reads:
\begin{equation}
    J = \underbrace{\frac{J_{pos} - J_{pos,min}}{J_{pos,max} - J_{pos,min}}}_{\tilde{J}_{pos}} + 
    \underbrace{\frac{J_{vel} - J_{vel,min}}{J_{vel,max} - J_{vel,min}}}_{\tilde{J}_{vel}}
\end{equation}
with each normalized component $\tilde{J}_{pos}$ and $\tilde{J}_{vel}$ bounded between zero (ideal, error-free case) and one (worst expected error).

To determine the normalization bounds, $J_{pos,min}$ was derived from a realistic error reference based on the 2024 Q2 Galileo Performance Report \citep{galileo-report}, which specifies 95\% confidence level 2D horizontal and vertical accuracies of $1.6\,\mathrm{m}$ and $2.4\,\mathrm{m}$, respectively. Assuming these values correspond approximately to $2\sigma$, this provides an estimate of the noise standard deviation $\sigma$ in a Gaussian white noise model. Applying this noise to the GT trajectory, we simulate a best-case, nominal error scenario for the navigation solution, thereby obtaining $J_{pos,min}$. The upper bound $J_{pos,max}$ is set by simulating a degraded scenario with noise standard deviations increased by a factor of five, representing a conservative worst-case.
Similarly, velocity normalization bounds are assigned based on empirical 95\% confidence intervals from \citep{velocity-accuracy}, which specify error thresholds of $0.05\,\mathrm{ms}^{-1}$ for the static case, $0.5\,\mathrm{ms}^{-1}$ for the rocket case, and $0.1\,\mathrm{ms}^{-1}$ for the LEO satellite case. Corresponding minimum $J_{vel,min}$ and maximum $J_{vel,max}$ velocity costs are computed by modeling measurement noise at nominal and degraded levels (five times the nominal standard deviation).

The proposed normalization strategy enables HyGO to effectively balance position and velocity accuracies as equally important optimization objectives within a single scalar cost, promoting solutions that achieve comprehensive navigation performance rather than optimization biased to one metric. This approach also ensures comparability and interpretability of cost values during optimization, simplifying convergence analysis and performance assessment.

%------------------- HyGO optimizer ----------------------
\subsection{HyGO: A Two-Stage Parametric Genetic Algorithm for GNSS Receiver Optimization} \label{ss:hygo}
The optimization landscape for GNSS receiver parameter tuning poses unconventional challenges, including nonlinear interdependencies, non-smoothness, collinearity, and the moderate-to-high dimensionality of the design space. To robustly navigate these issues, this study adopts the Hybrid Genetic Optimizer (HyGO), a binary-encoded genetic algorithm validated in prior high-dimensional control and engineering applications \citep{Robledo2025Hygo, rodrigo-turbulent}. While HyGO natively supports hybridization with direct search strategies, here a simplified two-stage refinement is implemented: an initial coarse search efficiently targets promising parameter regions, followed by a finer discretization to polish candidate solutions near global minima. This approach judiciously balances robustness, precision, and computational efficiency in an SDR context.

Genetic algorithms are well-suited for the highly multi-modal, noisy, and non-convex cost landscapes observed in GNSS applications. Their ability to efficiently traverse large, poorly characterized solution spaces without smoothness or gradient assumptions is critical in both simulation and field scenarios \citep{Robledo2025Hygo, rodrigo-turbulent}. HyGO’s binary encoding facilitates disciplined exploration, leveraging discrete representation to model both physical and algorithmic constraints on receiver parameters. Robustness to environmental error sources (multipath, atmospheric delays, equipment noise, orbital and clock biases) is a key advantage over gradient-based methods in GNSS receiver design \citep{gnss-errors}.

HyGO encodes the receiver controller parameters, defined in $\boldsymbol{\theta}$ (see previous section), into binary chromosomes representing discrete steps within parameter bounds. Each individual in the population corresponds to a particular parameter set, decoded for evaluation by the cost function $J$ (see Subsection \ref{ss:cost-function}). The optimization iteratively evolves the population across generations by applying genetic operators: elitism favors individuals with lower cost, crossover recombines parameter bit strings, and mutation introduces random variations. Through this evolutionary process, the population converges toward parameter sets minimizing the cost function, thus optimizing the receiver configuration \citep{genetic-definition}.

Informed by prior benchmarking and solver calibration, the selection of HyGO’s hyperparameters is summarized in \autoref{tab:hygo-settings}. The population is set to 100 individuals per generation, with A 10-generation limit; however, early termination is triggered if no cost improvement occurs in three consecutive generations, conserving computational resources. The rocket and LEO satellite cases are solved with seven controllers, while the static case includes eight due to integration time. Replication is disabled to maximize candidate diversity and computational efficiency. One elitist individual per generation ensures monotonic improvement, and both crossover and mutation probabilities are set to 0.5 to maintain a balanced search, preventing dominance of either exploitation or exploration. The mutation rate (0.06) injects variability without destabilizing convergence.
HyGO employs tournament-based selection for genetic operations, using the methodology and heuristics from \citep{Robledo2025Hygo}. A tournament size of 7 and selection probability of 1.0 maximizes competitive pressure and diversity in the selection process, balancing global exploration with local refinement, as substantiated by HyGO’s prior performance in complex tasks.
\begin{table}
    \centering
    \begin{tabular}{ll}
    \hline
    \textbf{Genetic Operator Parameter} & \textbf{Value} \\ \hline \hline
    Number of controllers         & 7/8     \\
    Individuals per generation    & 100   \\
    Number of generations         & 10    \\
    Replication probability       & 0   \\
    Crossover probability         & 0.5   \\
    Mutation probability          & 0.5   \\
    Elitism individuals           & 1     \\
    Mutation rate                 & 0.06   \\ 
    Tournament size               & 7      \\
    Tournament selection probability & 1    \\\hline
    \end{tabular}
    \caption{HyGO’s core genetic operator settings, including controller count (scenario-dependent), population size, generation ceiling, selection mechanics, and operator probabilities.}
    \label{tab:hygo-settings}
\end{table}

\begin{figure}
    \centering
    \includegraphics[width=0.99\textwidth]{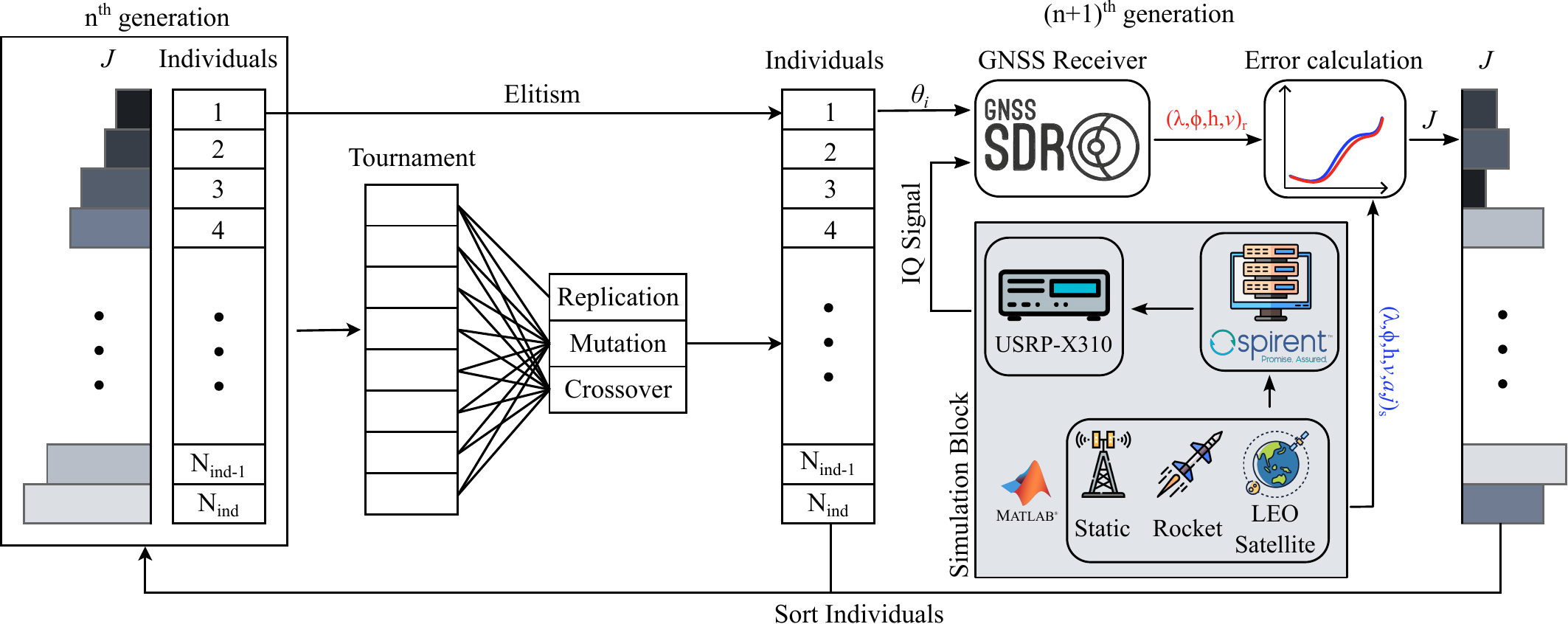}
    \caption{Schematic overview of the HyGO genetic algorithm optimization loop for GNSS receiver parameter tuning. The iterative workflow comprises population initialization (with Latin Hypercube Sampling), cost evaluation via hardware-in-the-loop simulation, tournament selection, elitism, and convergence. Operational scenarios (static, rocket, LEO) are modeled.}
    \label{hygo-optimization-loop}
\end{figure}
The schematic in \autoref{hygo-optimization-loop} encapsulates the full GA optimization cycle. An initial population of individuals, each encoding parameter vectors as binary strings, is evaluated in parallel through the SDR-based GNSS receiver chain, simulating varied operational scenarios (static, rocket, LEO satellite). Elitism retains the best solution across generations, while tournament selection drives eligible individuals into replication, crossover, and mutation pools. Resolution refinement is achieved via bit-length increases after convergence in the coarse phase, accelerating precise search near minimum solutions. Population initialization leverages Latin Hypercube Sampling (LHS) to uniformly sample the multi-dimensional parameter space, ensuring comprehensive initial coverage and improved convergence reliability \citep{lhs}. Parameter step sizes for controller settings are intentionally coarser (generations 1–5) to facilitate global search and are refined (generations 6–10) for local tuning. This two-phase resolution strategy substantially improves both escape from local minima and final solution precision. \autoref{tab:params-resolution-coarse-fine} details the discretization steps for coarse and fine generations, as well as chromosome length.
\begin{table}[H] 
    \centering
    \begin{tabular}{@{}lccccc|cccl@{}}
        \toprule
        \multicolumn{1}{c}{\textbf{Resolution}} & \textbf{\begin{tabular}[c]{@{}c@{}}PLL BW\\ {[}Hz{]}\end{tabular}} & \textbf{\begin{tabular}[c]{@{}c@{}}PLL BW \\ narrow \\ (\% PLL)\end{tabular}} & \textbf{\begin{tabular}[c]{@{}c@{}}DLL BW\\ {[}Hz{]} \end{tabular}} & \textbf{\begin{tabular}[c]{@{}c@{}}DLL BW \\ narrow \\ (\% DLL)\end{tabular}} & \textbf{\begin{tabular}[c]{@{}c@{}}FLL BW\\ {[}Hz{]}\end{tabular}} & \textbf{Static} & \textbf{Rocket} & \textbf{\begin{tabular}[c]{@{}c@{}}LEO \\ satellite\end{tabular}} & \textbf{\begin{tabular}[c]{@{}l@{}}Chromosome\\ length\end{tabular}} \\ \midrule
        Coarse & 5 & 5 & 3 & 5 & 3 & 30 & 27 & 27 & Coarse \\
        Fine & 1 & 1 & 1 & 1 & 1 & 39 & 36 & 36 & Fine \\ \bottomrule
    \end{tabular}
    \caption{Controller parameter discretization steps for coarse and fine resolution phases. Coarse steps support global search; fine steps facilitate local tuning. PLL and DLL filter orders are restricted to integer values. Chromosome length increases with finer resolution, enabling higher precision. Integration time is fixed at 1 ms except for static cases.}
    \label{tab:params-resolution-coarse-fine}
\end{table}

HyGO is implemented in Python, offering flexible GA configurations and supporting hybridization with local search strategies. While HyGO’s hybrid capabilities are reserved for future investigation, this study focuses on pure GA to maximize exploratory behavior in challenging, noisy landscapes.

%--------------------------------------------------------
%--------------------- RESULTS --------------------------
%--------------------------------------------------------
\section{Results} \label{results}
The performance of the Genetic Algorithm (GA) is analyzed for the three dynamic scenarios: static, rocket, and LEO satellite. After removing repeated individuals, a total of 990, 988, and 991 unique parameter combinations were evaluated for each case, respectively. 

\begin{figure}
    \centering
    \includegraphics[width=1\textwidth]{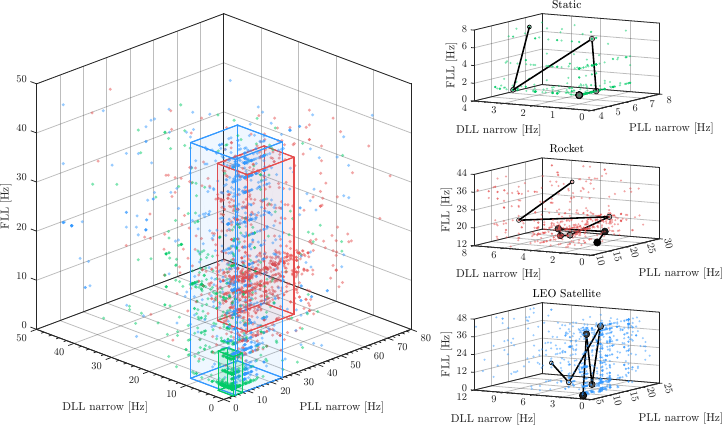}
    \caption{ Distribution of the three main GNSS tracking loop parameters: PLL narrow, DLL narrow, and FLL bandwidth. \textbf{Left}: 3D representation of the full population, illustrating how individuals are spread across the parameter space. \textbf{Right}: Zoomed-in views for each scenario (static, rocket, LEO satellite), where the best individual from each generation is marked with a grayscale dot whose size and intensity increase with the generation number. These best individuals are connected by \lcap{-}{black}, highlighting the trajectory of the optimization across generations. Individuals are represented as \sy{static}{o*} for static, \sy{rocket}{o*} for rocket, and \sy{satellite}{o*} for LEO satellite scenarios.}
    \label{pll_dll_fll}
\end{figure}

 The PLL and DLL narrow bandwidths, as well as the FLL bandwidth of the GNSS tracking loops once bit synchronization is performed, are represented in \autoref{pll_dll_fll}. 
Individuals appear to be more spread out at the beginning, as the initial population is evenly distributed across the search space using the Latin Hypercube Sampling (LHS) algorithm. As the GA evolves, different trends emerge for each scenario, and the individuals begin to cluster toward specific regions of the search space.

\begin{itemize}
    \item The static case quickly tends to minimize PLL, DLL, FLL bandwidth values, among the allowed ranges. 
    \item The rocket case opens up the bandwidth of the PLL, and shows that a FLL tracking loop is needed. 
    \item The LEO satellite case evolve towards a population with small PLL, and DLL bandwidths. The algorithm evaluated different FLL values, but finally tends to the smallest FLL bandwidth allowed. 
\end{itemize}

\begin{figure}[h]
    \centering
    \includegraphics[width=0.99\textwidth]{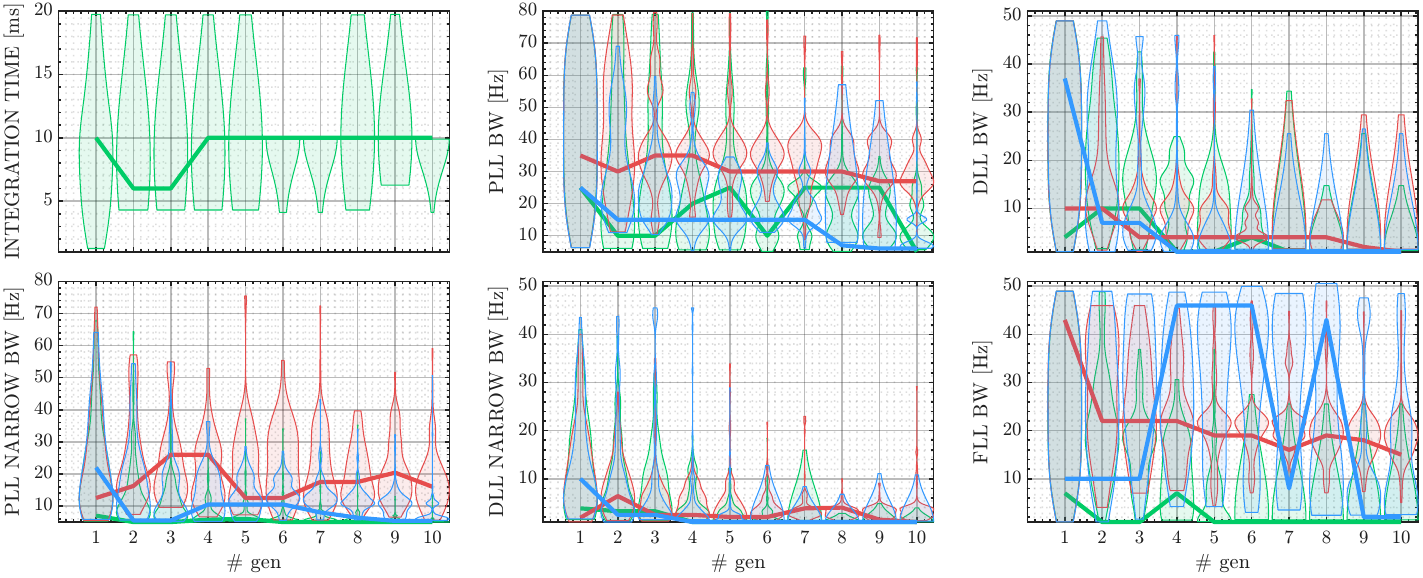}
    \caption{Representation of the parameters: integration time, PLL bandwidth, PLL narrow bandwidth, DLL bandwidth, DLL narrow bandwidth, and FLL bandwidth. The KDE distributions are represented by the colors \sq{static_violin} for the static case,\sq{rocket_violin} for the rocket case, and \sq{satellite_violin} for the LEO satellite case. The lines \lincW{static}, \lincW{rocket}, and \lincW{satellite} represent the generational evolution of each parameter for the best individual of each generation, respectively.}
    \label{param-evolve}
\end{figure}

\autoref{param-evolve} represents the evolution of all optimized variables, except the order of the filters. It shows a set of violin plots that illustrate how the distribution of parameter values evolves across generations. Each violin corresponds to one generation and displays the frequency with which different values appear in the population. The shape is based on a Kernel Density Estimation (KDE) using a Gaussian kernel, which smooths the data to show how densely different parameter values are populated. The wider regions indicate a higher concentration of individuals in that range, while the narrower regions reveal fewer explorations. This visualization highlights how the search narrows or shifts across generations as the GA converges. 

\begin{table}[H]
\begin{tabular}{ccccccccc}
\hline
\textbf{Case} & \textbf{\begin{tabular}[c]{@{}c@{}}Integration\\ Time {[}ms{]}\end{tabular}} & \textbf{\begin{tabular}[c]{@{}c@{}}PLL BW \\ {[}Hz{]}\end{tabular}} & \textbf{\begin{tabular}[c]{@{}c@{}}PLL BW \\ narrow \\ {[}Hz{]}\end{tabular}} & \textbf{\begin{tabular}[c]{@{}c@{}}PLL \\ filter order\end{tabular}} & \textbf{\begin{tabular}[c]{@{}c@{}}DLL BW \\ {[}Hz{]}\end{tabular}} & \textbf{\begin{tabular}[c]{@{}c@{}}DLL BW \\ narrow \\ {[}Hz{]}\end{tabular}} & \textbf{\begin{tabular}[c]{@{}c@{}}DLL \\ filter order\end{tabular}} & \textbf{\begin{tabular}[c]{@{}c@{}}FLL BW\\ {[}Hz{]}\end{tabular}} \\ \hline
Static        & 10                                                                           & 5                                                                   & 5                                                                             & 2                                                                    & 1                                                                   & 1                                                                             & 3                                                                    & 1                                                                  \\
Rocket        & 1                                                                            & 27                                                                  & 16                                                                            & 3                                                                    & 1                                                                   & 1                                                                             & 1                                                                    & 15                                                                 \\
LEO satellite & 1                                                                            & 6                                                                   & 5                                                                             & 3                                                                    & 1                                                                   & 1                                                                             & 1                                                                    & 1                                                                  \\ \hline
\end{tabular}
\caption{Optimum parameters found for each scenario}
\label{parameters_list}
\end{table}

\autoref{parameters_list} summarizes the parameters found for the best individual of the optimization for each case of analysis. The evolution of the tracking loop parameters across generations reveals distinct adaptation patterns for each case and parameter:
\begin{itemize}
\item  PLL narrow bandwidth: During the generations 1-5, values are widely distributed for the rocket and satellite cases. From generation 6 onward, both scenarios tend to converge. The rocket case concentrates the majority of the population between 10-30 Hz, while the LEO satellite reduces it between 5-10 Hz. The static case has a much faster convergence: at generation 4, has already all its individuals focused between 5-10 Hz and fixes all the population at 5 Hz from generation 6 onwards. 

\item PLL bandwidth: The rocket case bases its exploration on the range 20-50 Hz. The static case shows a more slow convergence of the exploration, until it finds its optimum at the area 5-10 Hz. The LEO satellite case explores a wide range of values between 5–40 Hz until generation 7, then opens up to search more widely in generations 8-9 and finally converges to a small area between 5-10 Hz.

\item DLL bandwidths (both normal and narrow): These parameters converge quickly in all three scenarios to the minimum value, indicating its role becomes less sensitive to dynamic variation. 

\item FLL bandwidth: The rocket case shows a clear convergence to a tight band between 15–25 Hz, reflecting the need to accommodate high dynamics early in flight. The LEO satellite case exhibits significant exploration across the full range but ultimately converges to the minimum. The static case shows minimal engagement with the FLL, underscoring its limited relevance in non-dynamic environments.

\item Integration time: After initial exploration, the static case fixes the value to 10 ms from generation 4 onward.

\item PLL filter order: The rocket and LEO satellite cases show a higher PLL filter order (3), compared to 2 for the static case. A higher-order filter increases the loop's ability to track rapid phase dynamics, such as acceleration and jerk, which take an important role in the rocket case. Although the LEO scenario involves high Doppler dynamics, it converges to the same PLL bandwidth (5 Hz) as the static case because the increased filter order (3) allows the loop to track rapid frequency variations without increasing bandwidth. In contrast, the static case benefits from the same narrow bandwidth for noise rejection, but does not require a higher filter order due to its low dynamic content.

\item DLL filter order: this parameter is mainly driven by the dynamics of the satellite–receiver distance. In the static scenario, since ranges vary slowly, so the value is 3. The dynamic scenarios require lower-order filters (1) to quickly adapt to rapid range changes.
\end{itemize}

\begin{figure}[H]
    \centering
    \includegraphics[width=1\textwidth]{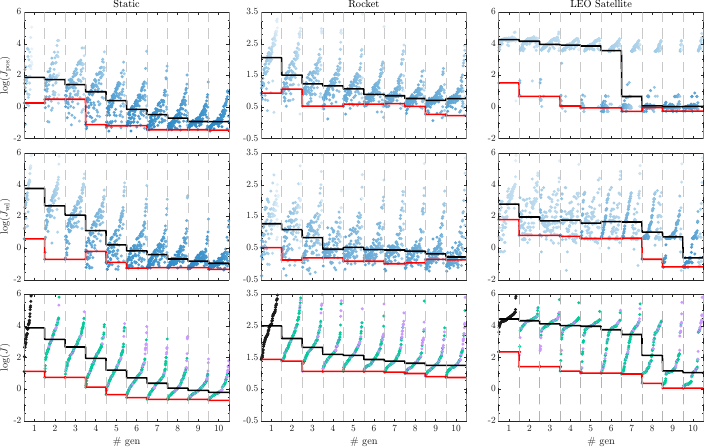}
    \caption{Evolution of the GA. Individuals are order from smaller to larger costs per generation. The blue dots in the $J_{pos}$ and $J_{vel}$ plots are color-coded based on their total $J$ cost, with lighter blue indicating higher cost and darker blue indicating lower cost. The evolution of the best individual value is represented by \lcap{-}{red}, while \lcap{-}{black} represents the median for each generation. Crossover operation is represented by \sy{crossover}{o*}, and mutation operation by \sy{mutation}{o*}. Elitism is highlighted in \sy{red}{o*}, and random  generation in \sy{black}{o*}}.
    \label{convergence}
\end{figure}

\autoref{convergence} represents on logarithmic scale how the cost of the individuals in terms of position and velocity evolves throughout the generations.

The GA converges progressively on each generation taking into account the contribution of the position ($J_{pos}$) and velocity ($J_{vel}$) errors, this can be seen in the trend of both the median and the best individual. The static case shows that on generation 4, the algorithm starts optimizing the position due to a mutation and grows slightly in velocity cost. From that point on, the velocity is also optimized. The rocket case shows the same behavior on generation 3, and finally converges to an optimum with enhaced position and velocity costs. The LEO satellite case shows two different areas that are explored from the generation 4. At generation 7, the algorithm finds that the new area is more suitable and is continually evolving around there. The difference between one area and the other is that the larger cost is associated to a loss of tracking at time 235 seconds.

The best individual is found at the end of the GA. The static and rocket cases find their best in the generation 10 and show an improvement of 83 $\%$ and 43 $\%$ with respect to the best of the initial generation. LEO satellite sets the best in  generation 9 and has an improvement of 89 $\%$. 

\begin{figure}[H]
    \centering
    \includegraphics[width=1\textwidth]{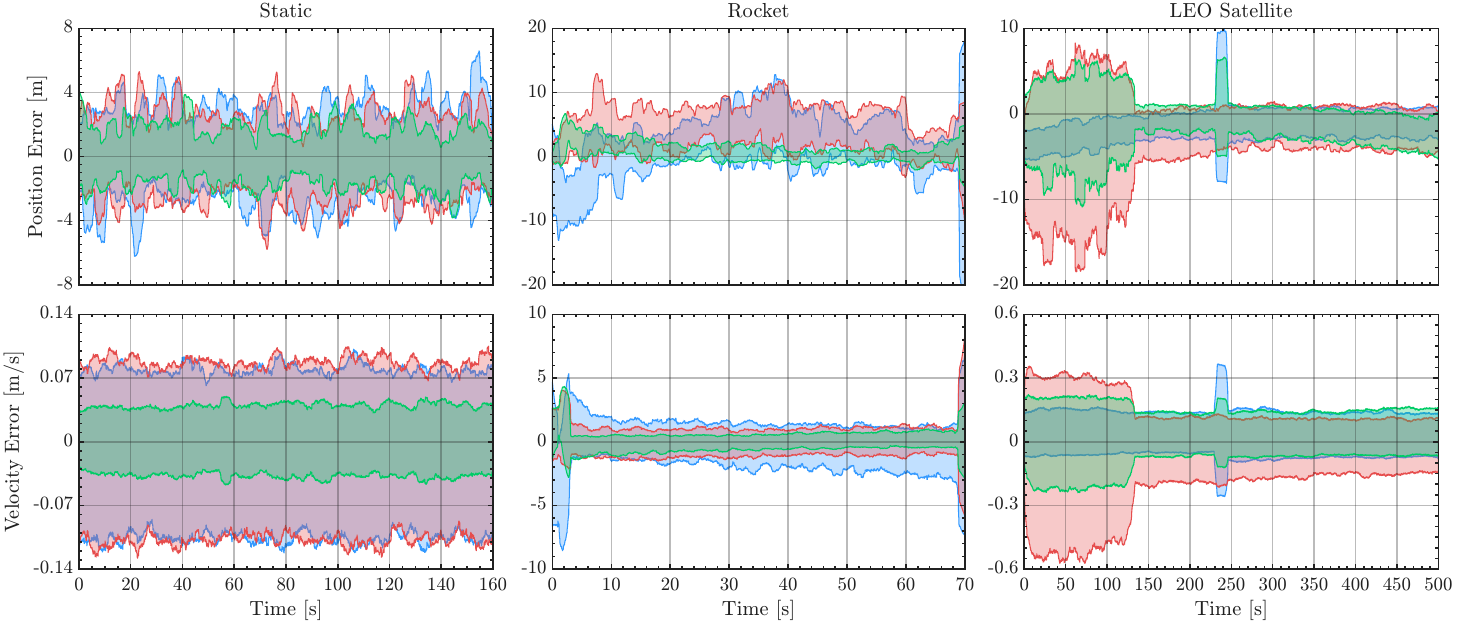}
    \caption{3$\sigma$ position and velocity errors. \sq{static_violin},\sq{rocket_violin}, and \sq{satellite_violin} colors represent x,y,z axes in ECEF reference frame, respectively.}
    \label{optimum-pos-vel}
\end{figure}

\autoref{optimum-pos-vel} contains the errors in position and velocity for the optimum cases for each scenario. The static user optimum establishes a maximum of 6 meters of error in position and 0.08 m/s in velocity. The rocket is able to estimate a valid navigation solution along the complete trajectory, despite dynamic adversities. At the maximum acceleration instant, the navigation solution exhibits an error of 12 m in position and 7 m/s in velocity, whereas the rest of the flight the velocity error is limited to 2 m/s. The satellite shows the first 120 seconds, where the PLL/DLL loops dominate and the errors can grow up to 12 m in position and 0.6 m/s in velocity, and then the error is reduced due to the dominance of the PLL/DLL narrow loops, where the position error is smaller than 5 meters and 0.2 m/s in velocity. The difference between the PLL/DLL loops and the PLL/DLL narrow loops is not visible in the static case, since they all have the smallest bandwidth value.

The optimization results reflect a well-established trade-off in GNSS receiver design involving the configuration of the predetection integration time, discriminator type, and loop filter bandwidth. To tolerate high dynamic stress, a receiver must adopt short integration times, frequency-based discriminators (FLL), and wideband loop filters. In contrast, to achieve high-accuracy carrier-phase measurements in low-dynamic conditions, long integration times, phase-based discriminators (PLLs), and narrow loop bandwidths are preferred. As discussed in \citep{theoretical-limit} Chapter 5.5, these requirements are inherently conflicting, and a compromise must be made depending on the application scenario.
This trade-off is clearly illustrated in the results obtained for the three evaluated scenarios:
\begin{itemize}
    \item In static scenarios, where Doppler and its derivatives are negligible, the optimizer prioritized measurement accuracy, resulting in longer integration times, PLL-only configurations, narrow bandwidths, and low-order PLL filters — all aligned with minimizing noise. These configurations minimized phase jitter by reducing thermal noise contribution, in line with the theory from Table 5.6 in \citep{theoretical-limit}, where first-order loops are noted as optimal in low-dynamic environments due to unconditional stability and low steady-state error.

    \item In contrast, the sounding rocket case illustrates the other extreme of the paradox: the optimizer favored shorter integration periods, FLL-assisted PLLs, and wide loop bandwidths to preserve lock under intense Doppler dynamics, sacrificing some measurement precision for tracking robustness.
    
    \item The LEO scenario, exhibiting smoother but non-negligible dynamics, produced a hybrid solution, maintaining relatively narrow bandwidths (like the static case) but increasing loop order to compensate for continuous Doppler rate variations — a nuanced resolution of the paradox.
\end{itemize}
In summary, these findings validate the practical design guideline that no single configuration is optimal across all motion regimes. A well-designed GNSS receiver must implement adaptive tracking strategies that resolve the inherent trade-off between measurement precision and dynamic resilience, as demonstrated through the automated optimization approach employed in this work.

\section{Conclusion} \label{conclusion}
This work has presented an automated methodology for optimizing GNSS receiver tracking loop configurations using a genetic algorithm, applied across three representative scenarios with distinct dynamic profiles: a static user, a LEO satellite, and a sounding rocket. These benchmarks span a broad spectrum of motion conditions relevant to GNSS operations, from stationary to highly accelerated dynamics, effectively addressing challenges across noise-limited and high-dynamic regimes.

The optimization was enabled by a fully software-defined architecture integrating realistic GNSS RF signal simulation, a hardware-in-the-loop RF front end, and the open-source GNSS-SDR receiver. This setup enabled the systematic evaluation of hundreds of receiver configurations without hardware reconfiguration, thus enabling large-scale, reproducible exploration of a complex multidimensional design space. The genetic algorithm employed a two-stage refinement strategy, combining a global coarse search with a high-resolution local tuning phase, enhancing convergence speed and solution precision.

Guided by a  cost function combining position and velocity errors, the optimization identified scenario-specific tracking loop parameters with clear physical interpretation. The static scenario converged to narrow PLL bandwidth of 5 Hz, FLL bandwidth of 1 Hz, and second-order PLL filter. The sounding rocket required 27 Hz narrow PLL bandwidth, 15 Hz FLL bandwidth, and third-order PLL filter. The LEO satellite balanced a narrow PLL bandwidth of 5 Hz, FLL bandwidth of 1 Hz, and third-order PLL filter. The results provided clear physical insights: narrow, low-order PLL loops were favoured under static, noise-limited conditions, whereas the sounding rocket required wider PLL bandwidths with higher-order filters to sustain lock under rapid Doppler changes. The LEO case, though characterized by high velocities, converged to narrower FLL bandwidth akin to the static scenario, consistent with the smooth Doppler evolution in orbital motion.

Performance improvements resulting from optimization were substantial, with maximum position and velocity errors reduced to $6~\mathrm{m}$ and $0.08~\mathrm{m/s}$ in the static case, 
$12~\mathrm{m}$ and $2.5~\mathrm{m/s}$ for the rocket, and $5~\mathrm{m}$ and $0.2~\mathrm{m/s}$ for the LEO scenario. These results underscore the efficacy and adaptability of genetic algorithm–based global optimization for tuning GNSS tracking loops, producing interpretable, physically consistent parameter sets tailored to diverse dynamic environments.

This study advances GNSS software-defined receiver design by demonstrating a robust, automated optimization framework that bridges signal simulation, hardware-in-the-loop experimentation, and evolutionary algorithms, providing a scalable landscape to realize optimized tracking architectures well matched to the demands of modern navigation challenges.

\section*{Acknowledgments}
This work has been partially supported by the Spanish Ministry of Science and Innovation under grant CPP2021-008648, funded by AEI/10.13039/501100011033 and co-financed by the European Union (NextGenerationEU/PRTR).
The authors would like to thank Carles Fernández and Javier Arribas for their support in reviewing results discussion and the GNSS architecture description, and Isaac Robledo for his guidance on HyGO hyperparameter configuration.

\section*{Data and Software Availability}
Simulation data used in this study will be available upon request. The HyGO framework is openly available to the research and engineering community under the MIT license, including source code, documentation, and ready-to-use examples. HyGO can be installed directly from the Python Package Index (PyPI) at \href{https://pypi.org/project/HYGO/}{\texttt{pypi.org/project/HYGO/}} by executing \texttt{pip install HYGO}. In addition, the full development repository is hosted on GitHub at \href{https://github.com/ipatazas/HYGO}{\texttt{github.com/ipatazas/HYGO}}, which includes a comprehensive suite of example scripts illustrating the use of HyGO for both parametric benchmark functions and control law optimisation—such as the stabilisation of the Landau oscillator.

The GNSS-SDR software-defined receiver is openly available at \href{https://gnss-sdr.org}{\texttt{gnss-sdr.org}}, including full source code, documentation, and community support resources. The GNSS-SDR configuration files, logs, and processing scripts employed in this work will also be made available upon reasonable request to foster reproducibility and facilitate further research on software-defined GNSS receiver optimisation.

\section*{Conflict of Interest}
The authors declare that they have no known competing financial interests or personal relationships that could have appeared to influence the work reported in this paper.

\bibliographystyle{abbrv} 
\bibliography{biblio.bib}

\end{document}

%% file: utils/defined_colors_and_commands.tex
\usepackage[normalem]{ulem}
\usepackage{color}
\usepackage{utils/mylines}
\usepackage{utils/myplots}
\newcommand{\sy}[2]{\mbox{(\kern-.25em\SymbolRGB[solid]{#1}{.8pt}{#2}{5pt}\kern-.25em)}}
\newcommand{\lsy}[3]{\mbox{(\kern-.1em\lineSymbolRGB{#1}{#2}{2pt}{#3}{4pt}\kern-.45em)}}
\newcommand{\lcap}[2]{~\,{\kern-1em\protect\mylcap{#1}{#2}}}
\newcommand{\lincW}[1]{(\,{\textcolor{#1}{\rule[1pt]{.4cm}{2pt}}}\,)}

\newcommand{\sq}[1]{(\textcolor{#1}{\rule[0pt]{.4cm}{8pt}})}

\definecolor{blue}{rgb}{0,0,1}
\definecolor{red}{rgb}{1,0,0}
\definecolor{black}{rgb}{0,0,0}
\definecolor{white}{rgb}{1,1,1}
\definecolor{grey}{RGB}{50,50,50}
\definecolor{redR}{RGB}{227, 47.3333, 39}
\definecolor{green}{RGB}{55, 160.3333, 85}
\definecolor{blue}{RGB}{55, 135, 192.3333}
\definecolor{yellow}{rgb}{0.9412, 0.7843, 0.0588}

\definecolor{static}{rgb}{0.0 0.8 0.4}
\definecolor{rocket}{rgb}{0.9 0.3 0.3}
\definecolor{satellite}{rgb}{0.2 0.6 1.0}
\definecolor{mutation}{RGB}{204, 153, 255}
\definecolor{crossover}{RGB}{0, 204, 153}

\colorlet{static_violin}{static!30!white}
\colorlet{rocket_violin}{rocket!30!white}
\colorlet{satellite_violin}{satellite!30!white}

%% file: MAIN_Arxiv.bbl
\begin{thebibliography}{10}

\bibitem{descent-vs-ga}
F.~Ahmad, N.~A.~M. Isa, M.~K. Osman, and Z.~Hussain.
\newblock Performance comparison of gradient descent and {G}enetic {A}lgorithm based {A}rtificial {N}eural {N}etworks training.
\newblock In {\em 10th International Conference on Intelligent Systems Design and Applications IEEE}, pages 604--609.

\bibitem{leosoops}
K.~W. Amir Allahvirdi-Zadeh, Ahmed El-Mowafy.
\newblock {D}oppler {P}ositioning {U}sing {M}ulti-{C}onstellation {LEO} {S}atellite {B}roadband {S}ignals as {S}ignals of {O}pportunity.
\newblock 72(2).

\bibitem{gnss-launchers}
B.~Braun, M.~Markgraf, and O.~Montenbruck.
\newblock {P}erformance analysis of {IMU}‑augmented {GNSS} tracking systems for space launch vehicles.
\newblock 8:117--133.

\bibitem{fll-justification}
C.~Cahn, D.~Leimer, C.~Marsh, F.~Huntowski, and G.~LaRue.
\newblock Software implementation of a {PN }spread spectrum {R}eceiver to accommodate dynamics.
\newblock 25(8):832--840.

\bibitem{rodrigo-turbulent2}
R.~Castellanos, G.~Y. Cornejo~Maceda, I.~de~la Fuente, B.~R. Noack, A.~Ianiro, and S.~Discetti.
\newblock Machine-learning flow control with few sensor feedback and measurement noise.
\newblock 34(4).

\bibitem{rodrigo-turbulent}
R.~Castellanos, A.~Ianiro, and S.~Discetti.
\newblock Genetically-inspired convective heat transfer enhancement in a turbulent boundary layer.
\newblock 230:120621.

\bibitem{particle-deep}
R.~Chaganti, A.~Mourade, V.~Ravi, N.~Vemprala, A.~Dua, and B.~Bhushan.
\newblock A particle swarm optimization and deep learning approach for intrusion detection system in internet of medical things.
\newblock 14(19):12828.

\bibitem{robust-tracking}
I.~Cortés, S.~Urquijo, M.~Overbeck, W.~Felber, L.~Agrotis, V.~Mayer, E.~Schönemann, and W.~Enderle.
\newblock {R}obust tracking strategy for modern {GNSS} receivers in sounding rockets.
\newblock In {\em 2022 10th Workshop on Satellite Navigation Technology (NAVITEC) IEEE}, pages 1--7.

\bibitem{fll_hydyn}
J.~T. Curran, G.~Lachapelle, and C.~C. Murphy.
\newblock Improving the design of frequency lock loops for {GNSS} receivers.
\newblock 48(1):850--868, 2012.

\bibitem{ga-gain-scheduling}
A.~S. Elkhatem and S.~N. Engin.
\newblock Enhancing performance and stability of gain-scheduling control system using evolutionary algorithms: A case study on transport aircraft.
\newblock 213:118859.

\bibitem{atc-genetic}
R.~S. F{\'e}lix~Patr{\'o}n, M.~Schindler, and R.~M. Botez.
\newblock Aircraft trajectories optimization by genetic algorithms to reduce flight cost using a dynamic weather model.
\newblock In {\em 15th AIAA Aviation Technology, Integration and Operations Conference}, page 2281, 06.

\bibitem{gnss-sdr}
C.~{Fern\'{a}ndez--Prades}, J.~Arribas, P.~Closas, C.~Avil\'{e}s, and L.~Esteve.
\newblock {GNSS-SDR}: An open source tool for researchers and developers.
\newblock In {\em Proceedings of the 24th International Technical Meeting of The Satellite Division of the Institute of Navigation (ION GNSS)}, pages 780--794, Portland, Oregon, Sept.

\bibitem{leognss}
H.~Ge, B.~Li, S.~Jia, L.~Nie, T.~Wu, Z.~Yang, J.~Shang, Y.~Zheng, and M.~Ge.
\newblock {LEO} enhanced global navigation satellite system ({L}e{GNSS}): progress, opportunities, and challenges.
\newblock 25(1):1--13.

\bibitem{gnss-leo}
E.~Gill, J.~Morton, P.~Axelrad, D.~M. Akos, M.~Centrella, and S.~Speretta.
\newblock Overview of space‑capable global navigation satellite systems receivers: heritage, status and the trend towards miniaturization.
\newblock 23(17):7648.

\bibitem{gnss-errors}
M.~S. Grewal, L.~R. Weill, and A.~P. Andrews.
\newblock {\em Global navigation satellite systems, inertial navigation and integration}.
\newblock John Wiley \& Sons.

\bibitem{gnss-sdr-escribano}
M.~Gómez, D.~Galindo, J.~Arribas, and C.~Fernández-Prades.
\newblock {GNSS} receiver developed on a {SDR} platform withstands high accelerations and speeds.
\newblock In {\em Proceedings of the 35th International Technical Meeting of the Satellite Division of The Institute of Navigation (ION GNSS+ 2022)}, pages 3333--3339, 9.

\bibitem{iridium}
N.~Jardak and Q.~Jault.
\newblock The potential of {LEO} satellite-based opportunistic navigation for high dynamic applications.
\newblock 22(7):2541.

\bibitem{genetic-definition}
W.~Jin, Z.~Hu, and C.~Chan.
\newblock A {G}enetic-{A}lgorithms-{B}ased {A}pproach for {P}rogramming {L}inear and {Q}uadratic {O}ptimization {P}roblems with {U}ncertainty.
\newblock 2013(1):272491, 01.

\bibitem{osnma}
J.~Juang, Y.~Chen, and C.~Chua.
\newblock {I}nter-{S}atellite and {I}nter-{R}eceiver {A}iding in the {V}erification of {OSNMA}.
\newblock In {\em Proceedings of the 36th International Technical Meeting of the Satellite Division of The Institute of Navigation (ION GNSS+ 2023)}, pages 483--494, Denver, Colorado. Institute of Navigation.

\bibitem{theoretical-limit}
E.~D. Kaplan and C.~J. Hegarty.
\newblock {\em Understanding {GPS}: Principles and Applications}.
\newblock Artech House, 2nd edition.

\bibitem{sdr-foundation}
R.~Krishnan, R.~G. Babu, S.~Kaviya, N.~P. Kumar, C.~Rahul, and S.~S. Raman.
\newblock Software defined radio ({SDR}) foundations, technology tradeoffs: A survey.
\newblock In {\em 2017 IEEE international conference on power, control, signals and instrumentation engineering (ICPCSI)}, pages 2677--2682. IEEE, 2017.

\bibitem{lhs}
M.~D. McKay, R.~J. Beckman, and W.~J. Conover.
\newblock A comparison of three methods for selecting values of input variables in the analysis of output from a computer code.
\newblock 42(1):55--61.

\bibitem{5dof}
{NATO Standardization Office (NSO)}.
\newblock {STANAG} 4355: {}the {M}odified {P}oint {M}ass and {F}ive {D}egrees of {F}reedom {T}rajectory {M}odel.
\newblock Technical report, 2009.
\newblock Accessed: 2025-06-01.

\bibitem{ins-calib}
X.~Niu, Y.~Li, H.~Zhang, Q.~Wang, and Y.~Ban.
\newblock Fast thermal calibration of low-grade inertial sensors and inertial measurement units.
\newblock {\em Sensors}, 13(9):12192--12217, 2013.

\bibitem{aero-genetic}
R.~Pe{\~n}a-Garc{\'\i}a, R.~D. Vel{\'a}zquez-S{\'a}nchez, C.~G{\'o}mez-Daza-Argumedo, J.~O. Escobedo-Alva, R.~Tapia-Herrera, and J.~A. Meda-Campa{\~n}a.
\newblock Physics-{B}ased {A}ircraft {D}ynamics {I}dentification {U}sing {G}enetic {A}lgorithms.
\newblock {\em Aerospace}, 11(2):142, 2024.

\bibitem{velocity-accuracy}
M.~Petovello.
\newblock How does a {GNSS} receiver estimate velocity?
\newblock {\em Inside GNSS}, pages 18--21, March/April 2015.
\newblock Accessed: 2025-06-01.

\bibitem{Robledo2025Hygo}
I.~Robledo, Y.~Li, G.~Cornejo~Maceda, and R.~Castellanos.
\newblock Fast and robust parametric and functional learning with {H}ybrid {G}enetic {O}ptimisation ({H}y{GO}).
\newblock {\em Arxiv}, 2025.

\bibitem{pll-assisted-fll}
P.~A. Roncagliolo, C.~E. De~Blasis, and C.~H. Muravchik.
\newblock {GPS} {D}igital {T}racking {L}oops {D}esign for {H}igh {D}ynamic {L}aunching {V}ehicles.
\newblock pages 41--45, 2006.

\bibitem{roncaglio1}
P.~A. Roncagliolo, J.~G. Garc{\'i}a, and C.~H. Muravchik.
\newblock {O}ptimized {C}arrier {T}racking {L}oop {D}esign for {R}eal-{T}ime {H}igh-{D}ynamics {GNSS} {R}eceivers.
\newblock 2012(1):651039.

\bibitem{sounding-rockets}
G.~Seibert and B.~T. Battrick.
\newblock {\em The history of sounding rockets and their contribution to {E}uropean space research}.
\newblock ESA Publications division Noordwijk.

\bibitem{radar-rockets}
L.~W.~T. Silva, V.~F. Barros, and S.~G. Silva.
\newblock Genetic algorithm with maximum-minimum crossover ({GA-MMC}) applied in optimization of radiation pattern control of hased-array radars for rocket tracking systems.
\newblock 14(8):15113--15141.

\bibitem{gnss-vs-ins}
D.~H. Titterton and J.~L. Weston.
\newblock {\em Strapdown Inertial Navigation Technology}, volume~17.
\newblock IET.

\bibitem{galileo-report}
{U.S. Federal Aviation Administration, William J. Hughes Technical Center}.
\newblock Galileo {O}pen {S}ervice {P}erformance {A}nalysis {R}eport 2024 {Q2}.
\newblock Technical report.
\newblock Accessed: 2025-06-01.

\bibitem{gnss-uav}
C.~Vogelsang et~al.
\newblock How {G}ood is a {T}actical-grade {GNSS+ INS} ({MEMS} and {FOG}) in a 20-m {B}athymetric {S}urvey?
\newblock 23(2):754.

\bibitem{optimization-methods}
E.~Zakharova and I.~Minashina.
\newblock Review of multidimensional optimization methods.
\newblock 60(6):625--636, 6.

\bibitem{steepest-ascent-method}
Q.~Zeng, W.~Qiu, J.~Liu, R.~Xu, J.~Shi, and Y.~Sun.
\newblock A high dynamics algorithm based on steepest ascent method for {GNSS} receiver.
\newblock {\em Chinese Journal of Aeronautics}, 34(12):177--186, 03 2021.

\end{thebibliography}
